\documentclass{elsart}
\usepackage{amssymb}
\usepackage{amsmath}
\usepackage{graphicx}
\newtheorem{theorem}{Theorem}[section]
\newtheorem{defit}{Definition}[section]

\begin{document}

\begin{frontmatter}

\title{Stationary modulated-amplitude waves in the 
	1-$D$ complex Ginzburg-Landau equation}
\author{Yueheng Lan, Nicolas Garnier and Predrag Cvitanovi\'c}
\address{Center for Nonlinear Science, School of Physics,
	Georgia Institute of Technology \\
	837 State Street, Atlanta, GA 30332-0430, USA}
\date{July 29th 2002}
\maketitle

\begin{abstract}

We reformulate the one-dimensional complex Ginzburg-Landau equation as a
fourth order ordinary differential equation in order to find stationary
spatially-periodic solutions. Using this formalism, we prove the
existence and stability of stationary modulated-amplitude wave
solutions. Approximate analytic expressions and a comparison with
numerics are given.

\end{abstract}


\end{frontmatter}

\section*{Introduction}

The cubic complex Ginzburg-Landau equation (CGLe) is a generic amplitude
equation describing Hopf bifurcation in spatially extended systems,
{i.e.}, $I_o$ systems~\cite{patt}, with reflection
symmetry~\cite{sup,sub,amp}. It is of great interest due to its
genericity and applications to onset of wave pattern-forming
instabilities~\cite{patt} in various physical systems such as fluid
dynamics, optics, chemistry and biology. It exhibits rich dynamics and
has become a paradigm for the transition to spatio-temporal chaos.

We consider the one-dimensional CGLe for the complex amplitude field
$A(x,t)$:
\begin{eqnarray} 
A_t=\mu A+(1+i\alpha)A_{xx}-(1+i\beta)|A|^2 A 
\label{cgl} 
\end{eqnarray} 
where $A(x,t):\mathbb{R}^2 \mapsto \mathbb{C}$, and $\mu,\alpha,\beta
\in \mathbb{R}$, $x \in {\mathcal D}$. $\mathcal D$ is the spatial
domain on which the equation is defined. Interesting domains for us are
either the whole real axis or a finite box of length $L$ with periodic
boundary conditions. $\mu$ is the control parameter. Only $\mu>0$ is
considered because we study the supercritical Ginzburg-Landau equation;
one could set $\mu=1$ by appropriate rescaling of the time, space and
amplitude, but we keep it as a parameter for closer connection with
experimental results and previous literature. Coefficients $\alpha$ and
$\beta$ parametrize the linear and nonlinear dispersion. 

If both $\alpha$ and $\beta$ are set to $0$, we recover the real
Ginzburg-Landau equation (RGLe) in which only the diffusion term and the
stabilizing cubic term compete with each other and the linear term. A
Lyapunov functional exists in that case~\cite{patt} and the RGLe behaves
like a gradient system. Another limit ---~the nonlinear Schr\"{o}dinger
equation~--- results from setting $\alpha,\beta \rightarrow \infty$; we
then have an integrable nonlinear PDE. For parameter values in the
intermediate range, long-time behavior of the CGLe can vary from
stationary to periodic and to spatiotemporal chaos~\cite{Shraiman:92}.
In this paper, we concentrate on the stationary solutions of the CGLe in
a finite box of length $L$ with periodic boundary conditions, and the
case $\alpha \neq \beta$. Stationary solutions are the simplest
non-trivial solutions, related to propagating solutions by an
appropriate change of frame of reference $(x,t) \mapsto (x-vt,t)$ with
fixed $v \in \mathbb{R}$. 

Searching for coherent structures allows one to reduce a partial
differential equation into an ordinary one, and such solutions of the
CGLe are believed to be extremely important in many regimes, including
the spatiotemporal chaos~\cite{hohen}. Recently, numerical integrations
of the CGLe have focused on a class of solutions called
modulated-amplitude waves (MAWs) and their role in the nonlinear
evolution of the Eckhaus instability of initially homogeneous plane
waves~\cite{maw,defect}. 

MAWs can bifurcate from the trivial solution $A=0$ (case I) or plane
wave solutions of zero wavenumber (case II). Analytical aspects of
modulated solutions of the CGLe have been addressed by Newton and
Sirovich who have applied a perturbation analysis to study the
bifurcation in case II~\cite{quarter1}, and discussed the secondary
bifurcation of those MAWs~\cite{quarter2}. Tak\'a\v{c}~\cite{pt} proved
the existence of MAW solutions using a standard bifurcation analysis in
the infinite-dimensional phase space of the CGLe, in both cases I and
II, together with a stability analysis in case I by means of the center
manifold theorem.

In this article we reformulate the CGLe equation assuming a coherent
structure form for the solutions, and obtain a fourth-order ordinary
differential equation (ODE) with a consistency condition. This form is
algebraically convenient, because the deduced system of four first-order
ODEs contains only quadratic non-linearity. In the Benjamin-Feir-Newell
regime, where plane waves solutions are always unstable, we give a proof
of existence of MAWs in both case I and II using our ODE. For weak
perturbations in case I or II, we write approximate analytic solutions
in the ODE phase space. Coming back to the full CGLe, we then prove the
stability of those MAWs in a finite box in case II, and prove that the
bifurcation is supercritical, as suggested by recent numerical
work~\cite{maw}.

In the next section, we discuss symmetries and solutions of the CGLe. In
section~\ref{sec:ODE} we transform the steady CGLe for MAWs into an
equivalent ODE, and give the sufficient condition to identify the
solutions of these two equations. In section~\ref{sec:existence} this
ODE is used to construct a 4-$D$ dynamical system and prove the existence
of symmetric stationary solutions of the CGLe in the two cases I and II.
In section~\ref{sec:approximate} the approximate analytic form of the
solutions is given and compared to numerical calculations, and the
stability of MAWs in case II is proved. Several theorems needed in the
proofs are reproduced in appendix~\ref{sec:theorems}.

\section{Basic properties of the CGLe}

\subsection{Symmetries}

The equation~(\ref{cgl}) is invariant under temporal and spatial
translations. Moreover, it is invariant under a global gauge
transformation $A\rightarrow A \exp({i\phi})$, where $\phi \in \mathbb{R}$,
and under $x \rightarrow -x$ reflection. As a consequence, it preserves
parity of A, i.e., if $A(-x,0)=\pm A(x,0) $, then $A(-x,t)=\pm A(x,t)$
for any later time $t>0$. If $A(x,t)$ has no parity, then $A(-x,t)$
gives another solution.

%

\subsection{Stokes solutions and their stability}

\begin{figure}\begin{center}
\includegraphics[width=8cm]{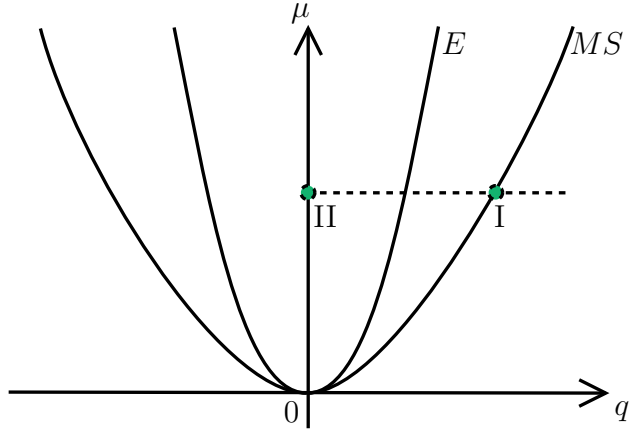}
\setlength\unitlength{1cm}
\begin{picture}(0,5)
\put(-4.5,5.5){$\mu$}
\put(-0.2,0.2){$q$}
\put(-0.8,5  ){$MS$}
\put(-2.5,5  ){$E$}
\put(-4.6,0.1){0}
\put(-4.2,2.7){II}
\put(-1.8,2.7){I}
\end{picture}
\end{center}
\caption{Marginal stability curve (MS) and Eckhaus instability 
	curve (E) defining regions where plane waves solutions 
	exist in the CGLe (inside (MS)), and are stable when $1+\alpha\beta>0$
	(inside (E)). Case I corresponds to the wave of maximal
	possible wavenumber and Case II to the wave with $q=0$ and 
	the maximal amplitude.}
\label{fig:diagram}
\end{figure}

The global phase invariance implies that the CGLe has nonlinear plane
wave solutions of form
\begin{equation}
A(x,t)=R_0 \exp ({i(qx-\omega t)}) \,,
\label{eq:Stokes}
\end{equation}
where $R_0^2=\mu-q^2$ is the amplitude squared, $\omega=\mu \beta +
(\alpha-\beta)q^2$ is the frequency, and $q \in \mathbb{R}, q^2 \leq
\mu$ is the wavenumber. They are called Stokes solutions~\cite{WoCGLe}
and are parametrized by the wavenumber $q$. The two limit cases of
interest to us are highlighted on figure~\ref{fig:diagram}: a plane wave
of wavenumber $\mu^{1/2}$ and of vanishing amplitude (case I), and the
wave with zero wavenumber and maximum amplitude $\mu$ (case II). In case
II, the solution oscillates uniformly in time; we call it the
homogeneously oscillating state (HOS).

For the infinite system, the Benjamin-Feir-Newell~\cite{Newell:74}
criterion states that all plane wave solutions are unstable with respect
to long wavelength perturbations (i.e., of wavenumber $k \rightarrow 0$)
if $1+\alpha \beta<0$. If $1+\alpha \beta>0$, we have to consider the
Eckhaus instability criterion; only a band of wavenumbers are stable
against long wavelength perturbations (figure~\ref{fig:diagram}):

\begin{equation}
q^2 < q_{\rm E}^2 \equiv 
\frac{(1+\alpha\beta) \mu}{ 3 + \alpha \beta + 2\beta^2} \,.
\label{eq:Eckhaus}
\end{equation}

For a finite periodic system the wavenumbers for both the original
states and the perturbations are quantized. These criteria have been
reexamined by Matkowsky and Volpert using linear stability
analysis~\cite{stp}.

\subsection{Coherent structures and MAWs}
\label{sec:intro:MAWs}

Coherent structures play a very important role in the study of pattern
formation and dynamical properties of the CGLe~\cite{hohen}. They are
uniformly propagating structures of the form
\[
A(x,t)=R(x-vt)\exp({i\phi (x-vt)}e^{-i\omega t})
\]
which can be expressed as solutions of a 3-$D$ nonlinear dynamical system
obtained by substituting the above ansatz into the CGLe. There are two
free parameters: the frequency $\omega$ and the group velocity $v$.

The fixed points of the 3-$D$ system are the plane waves described in
the previous section. The homoclinic~\cite{build} and
heteroclinic~\cite{hohen} connections between the fixed points
correspond to localized coherent structures. The Nozaki-Bekki
solutions~\cite{bk} belong to this category; they connect asymptotic
plane waves with different wavenumbers. In numerical simulations in
large domains, nearly coherent structures are frequently observed in
chaotic regimes, thus suggesting those objects are also relevant to
spatiotemporally chaotic dynamics.

Recent numerical studies reveal another kind of coherent structure:
modulated amplitude waves (MAWs) for the CGLe~\cite{maw}. They
correspond to limit cycles of the 3-$D$ nonlinear system. When $v=0$,
MAWs are stationary. The formation of MAWs is the first instability
encountered when a plane wave state crosses the Eckhaus or Benjamin-Feir
stability line. The MAW structure is frequently observed in
experiments~\cite{sup,garnier} and considered as a key to interpretation
of patterns and bifurcations exhibited during the system's transition to
spatio-temporal chaos~\cite{defect}. Traveling MAWs have been observed
in numerical simulations of the CGLe in periodic boxes, with parameter
$q$ between 0 and $\mu^{1/2}$, i.e., in between cases I and II; we are
interested here only in stationary MAWs that appear either in case I or
case II.

In this paper, we propose a new real-valued ODE to describe steady
solutions of the CGLe. A 4-$D$ dynamical system derived from this ODE
enables us to apply the successive approximation method~\cite{jhale},
to prove the existence of stationary MAWs and to give the analytical 
form of the approximate solutions in both case I and case II. Numerical
integrations of the exact CGLe are then compared to the approximate
analytic result. Furthermore, we show non-analyticity at discrete points
of solutions in case I, and prove the stability of the MAWs in case II.
Some theorems needed in our proof are reproduced in the
appendix~\ref{sec:theorems}. In what follows, $\mathrm{diag}(\cdots)$
denotes a (block) diagonal matrix and $\mathrm{col}(\cdots)$ a column
vector.

\section{Stationary case}
\label{sec:ODE}

Since we are only interested in the steady solutions of the CGLe, we
substitute the ansatz
\begin{equation}
 A(x,t)=R(x) \exp({i\phi(x)-i\omega t}), 
	\qquad (R,\phi) \in \mathbb{R}^2
\label{eq:antsatz}
\end{equation}
into (\ref{cgl}). We then have    
\begin{eqnarray}
(1+\alpha^2)G_x & = K & \equiv (\beta-\alpha) R^4
	-(\omega-\mu \alpha)R^2  \label{or1}\\ 
(1+\alpha^2)G^2 & = M & \equiv (1+\alpha^2)R^3 R_{xx}
	+(\alpha \omega +\mu)R^4 -(1+\alpha \beta) R^6  \,.  \label{or2} 
\end{eqnarray}
where $G \equiv \phi_x R^2$ is reminiscent of ``angular momentum''. Note
that if $\alpha=\beta$, this ``angular momentum'' is conserved ---~it is
constant in space~--- provided that $\omega=\mu \alpha$. In that case,
(\ref{or2}) can be solved in terms of elliptic functions~\cite{ellip}.
We will only consider the case $\alpha \neq \beta$ in the following.
Equations (\ref{or1}) and (\ref{or2}) are invariant under $(G,x)
\rightarrow (-G,-x)$. Note that for plane waves, $K=0$ and $G$ is a
constant. If $K$ is not always zero, differentiating (\ref{or2}) and
dividing the result by (\ref{or1}) gives
\begin{equation}
	2G = M_x/K \,,  \label{dg}
\end{equation} 
and by (\ref{or2})
\begin{eqnarray}
	M = \frac{1+\alpha^2}{4}\frac{M_x^2}{K^2} \,.\label{sqr}
\end{eqnarray} 

Furthermore, we can factorize $R^2$ from $M_x$ and $K$ and write
$M_x = R^2 N $ and $K=R^2 P$, where
\begin{eqnarray}
	N & \equiv & (1+\alpha^2)\frac{1}{2}(R^2)_{xxx}+(\alpha
		\omega+\mu)2(R^2)_x -(1+\alpha \beta)3R^2(R^2)_x 	
							\nonumber\\ 
	P & \equiv & (\beta-\alpha)R^2-(\omega-\mu \alpha) \,.		
							\label{eq:introP}
\end{eqnarray}
The last relation can be used to express $R^2$ in terms of $P$:
\begin{eqnarray}
	R^2=\frac{(\omega-\mu \alpha)+P}{\beta-\alpha}=e+dP
	= R_0^2 + \frac{P}{\beta-\alpha} \,,		\label{dR}
\end{eqnarray}
where $d\equiv 1/(\beta-\alpha)$ and 
$e \equiv {(\omega-\mu \alpha)}/{(\beta-\alpha)}$.

Note that $e=R_0^2$ is the square of the homogeneous amplitude
$R_0(q,\omega)$ of the Stokes plane wave solution (\ref{eq:Stokes}) of
frequency $\omega$ and wavevector $q(\omega)$. $P$ then appears as the
modulation of the amplitude squared with respect to the Stokes solution,
and so it is an appropriate variable to describe a MAW.

Substituting $K$ and $M_x$ into (\ref{sqr}), we have
\begin{equation}
	\frac{1+\alpha^2}{4} \frac{N^2}{P^2}=M \,. \label{or3}
\end{equation} 
If $P \neq 0$ (\ref{or3}) is equivalent to (\ref{or1}) and
(\ref{or2}). It is easy to check that if we regard
(\ref{dg}) as a definition of $G$, and use $K,M,N,P$ expressed in terms
of $R$, equation (\ref{or1}) and (\ref{or2}) will be recovered as a
result of (\ref{sqr}) and (\ref{or3}). Differentiating both sides of
(\ref{or3}) results in
\begin{equation}
	\frac{1+\alpha^2}{2}(PN_x-NP_x)  =  R^2 P^3 \,.		\label{rd}
\end{equation}

In this step we have extended the solution set of (\ref{or3}), because as 
we integrate (\ref{rd}) back, we get
\begin{equation}
	\frac{1+\alpha^2}{4} \frac{N^2}{P^2}= M + {\mathcal C} \,, 	\label{mrd}
\end{equation} 
where ${\mathcal C}$ is an integration constant. Only when ${\mathcal
C}=0$, a solution of (\ref{rd}) is a solution of (\ref{or3}). For this
reason, when obtaining solutions of (\ref{rd}), we have to check the
{\em consistency condition}
\begin{equation}
	\frac{1+\alpha^2}{4} \frac{N^2}{P^2}-M=0 		\label{cc1}
\end{equation}  
to make sure that we have a solution of (\ref{or3}), thus a solution of
(\ref{or1}) and (\ref{or2}). Note that if $K$ vanishes we have to go
back to (\ref{or1}) and (\ref{or2}), since in that case (\ref{or3}) is
not well defined. Let us rewrite $N$ in terms of $P$:
\begin{eqnarray}
N= \displaystyle \frac{2}{1+\alpha^2} \left(aP_{xxx}+bP_x+cPP_x \right)	\,, \label{dC}
\end{eqnarray}
where $a,b,c$ are constants
\begin{eqnarray}
a & \equiv & \frac{(1+\alpha^2)^2}{4(\beta-\alpha)} \nonumber \\
b & \equiv & \frac{1+\alpha^2}{2}\left(\frac{2(\alpha \omega+\mu)}{\beta-\alpha}-
\frac{3(1+\alpha \beta)(\omega-\mu \alpha) }{(\beta-\alpha)^2}\right) \label{eq:abcdefs} \\
c & \equiv & -\frac{3(1+\alpha \beta)(1+\alpha^2)}{2(\beta-\alpha)^2} \,. \nonumber
\end{eqnarray}
After some algebra (here relegated to appendix~\ref{sec:algebra}),
we get an equation for $P$ only:
\begin{equation}
\left(\frac{\tilde{M}_x}{P}\right)_x = \frac{\lambda}{a}\tilde{M} + kP \,,
\qquad \tilde{M} \equiv \lambda P_{xx}+dP^2+\tilde{e}P \,.
\label{ode1}
\end{equation} 

$\lambda$ is a fixed real constant that depends on $\alpha$ and $\beta$
only, and that takes two different values given in
appendix~\ref{sec:algebra}. $\lambda$ is a transient variable used in
the proof and derivation but our solutions to the CGLe do not depend on
$\lambda$ and do not distinguish the two values of $\lambda$ (see
section~\ref{sec:approximate}). $\tilde{e}$ and $k$ are real parameters
introduced as $\tilde{e}+\frac{a}{\lambda}k=e$. So (\ref{ode1}) has two
free parameters: $\omega$, introduced by the ansatz~(\ref{eq:antsatz})
as the carrier frequency of the solution, and $k$. the {\em consistency
condition} (\ref{cc1}) fixes one parameter.

\section{4-$D$ dynamical system and the existence of periodic solutions}
\label{sec:existence}

Let us take $\tau$ as the spatial variable, $P=P(\tau)$ in
(\ref{eq:introP}), and rewrite (\ref{ode1}) as a system of first order
equations in $\tau$. With $\tilde{N} = {\tilde{M}_\tau}/{P}$ and
$Q=P_\tau$, from (\ref{ode1}) we have
\begin{equation}
\left\{\begin{array}{rcl}
       \dot{\tilde{M}} &=& \tilde{N}P \\
       \dot{\tilde{N}} &=& \frac{\lambda}{a}\tilde{M}+kP \\
       \dot{P}         &=& Q \\
       \dot{Q}         &=& \frac{1}{\lambda}(\tilde{M}-dP^2-\tilde{e}P)  
       \end{array}\right. \,,
\label{eq:4d}
\end{equation}
where the dot represents the derivation with respect to the spatial
variable $\tau$.

It is easy to check that $P=0$ is a solution of the original equations
(\ref{or1}) and (\ref{or2}), corresponding to the plane wave solution of
the CGLe with frequency $\omega$. We will study the behavior near $P=0$
and prove the existence of periodic solutions for small $P$. In the CGLe,
this corresponds to a weakly modulated amplitude wave which bifurcates
from a plane wave solution. If $P \sim \epsilon$, where $\epsilon$ is a
small parameter, so are $\tilde{M},\tilde{N},Q$ by their
definitions. Write
\begin{equation*}
(\tilde{M}, \tilde{N}, P, Q)
= (\epsilon x, \epsilon y, \epsilon z, \epsilon w)
\end{equation*}
and set $k=k_1+\epsilon k_2$. Substituting these into the 4-$D$ system, we have
\[
\frac{d}{d\tau}\left(\begin{array}{c}
                    x \\ y \\ z \\ w
                  \end{array} \right)=A\left(\begin{array}{c}
        x \\ y \\ z \\ w
      \end{array} \right)+\epsilon \left(\begin{array}{c}
                                         y\,z \\ k_2\, z \\0\\ - \frac{d}{\lambda}z^2 
                                         \end{array} \right),\mbox{ where }
A=\left(\begin{array}{cccc}
                                             0 & 0 & 0                  & 0 \\
                             \frac{\lambda}{a} & 0 & k_1                & 0 \\
                                             0 & 0 & 0                  & 1 \\
                             \frac{1}{\lambda} & 0 & \frac{-\tilde{e}}{\lambda} & 0
                                            \end{array}\right). 
\]

The linear part $A$ describes the behavior of the system in the
neighborhood of the trivial fixed point $(0,0,0,0)$. Note that the
system is invariant under $(t,y,w) \rightarrow -(t,y,w)$. We use this
property to simplify our analysis. Moreover, this system defines an
incompressible flow since $\nabla \cdot \vec{X}=0$, where
$\vec{X}=(x,y,z,w)$. It follows from (\ref{mrd}) that the system
has one integration constant ${\mathcal C}$. This constant induces a
foliation of the phase space into three-dimensional manifolds. Physical
solutions, i.e., the solutions of the original CGLe, are restricted to
${\mathcal C}=0$, the manifold that satisfies the {\em consistency
condition} (\ref{cc1}). 

These properties strongly restrict the possible distribution of
eigenvalues of $A$. We restrict our analysis to the case
${\tilde{e}}/{\lambda}>0$, then $A$ has eigenvalues
$\left\{ 0, 0, i \omega_1, -i \omega_1 \right\}$ with $\omega_1 =
\sqrt{{\tilde{e}}/{\lambda}}$. In that case, periodic solutions or
MAWs can exist as we will prove in the following. 
The evolution of the system along either
of the two degenerate eigenvalue 0 directions respects the constant
${\mathcal C}$ foliation: if the solution is on a constant ${\mathcal
C}$ manifold at initial time, it remains there for any later time.

We now discuss the condition $\tilde{e}/\lambda>0$ in terms of an
instability of the underlying plane wave. We can rewrite
$\tilde{e}/\lambda$ using (\ref{eq:abcdefs}) and (\ref{condi}). Assuming
that the solution we are searching for is close to a plane wave, we can
use the wavenumber $q$ instead of the frequency $\omega$, using the
dispersion relation (\ref{eq:Stokes}) for plane waves:
\begin{eqnarray}
\frac{\tilde{e}}{\lambda} = \frac{b}{a} 
	& = & \frac{2}{1+\alpha^2}\left[2(\alpha \omega +\mu)
	 	 -\frac{3(1+\alpha \beta)(\omega-\mu \alpha)}
		{\beta-\alpha}\right]			\label{domega1} \\
	& = & \frac{2}{1+\alpha^2}\left[ (3+\alpha\beta+2\alpha^2)q^2
	  	-(1+\alpha\beta)\mu \right] \,. \nonumber 
\end{eqnarray}
If we write
\begin{equation}
q_{\rm M}^2 \equiv \frac{(1+\alpha\beta) \mu}{ 3 + \alpha \beta + 2\alpha^2} \,,
\label{eq:defqm}
\end{equation}
we have
\begin{eqnarray}
\frac{\tilde{e}}{\lambda} > 0 & \Leftrightarrow & 
\left| \begin{array}{rl}
q^2 > q_{\rm M}^2 
	& \quad {\rm if} \quad (1+\alpha\beta)>0 \\
q^2 < q_{\rm M}^2
 	& \quad {\rm if} \quad (1+\alpha\beta)<-2(1+\alpha^2) \\
\forall q \in [-\sqrt{\mu}, \sqrt{\mu}]
	& \quad {\rm if} \quad -2(1+\alpha^2) < 1+\alpha\beta < 0
\end{array} \right. .
\label{eq:positivity}
\end{eqnarray}

\begin{figure}
\begin{center}
\includegraphics[width=8cm]{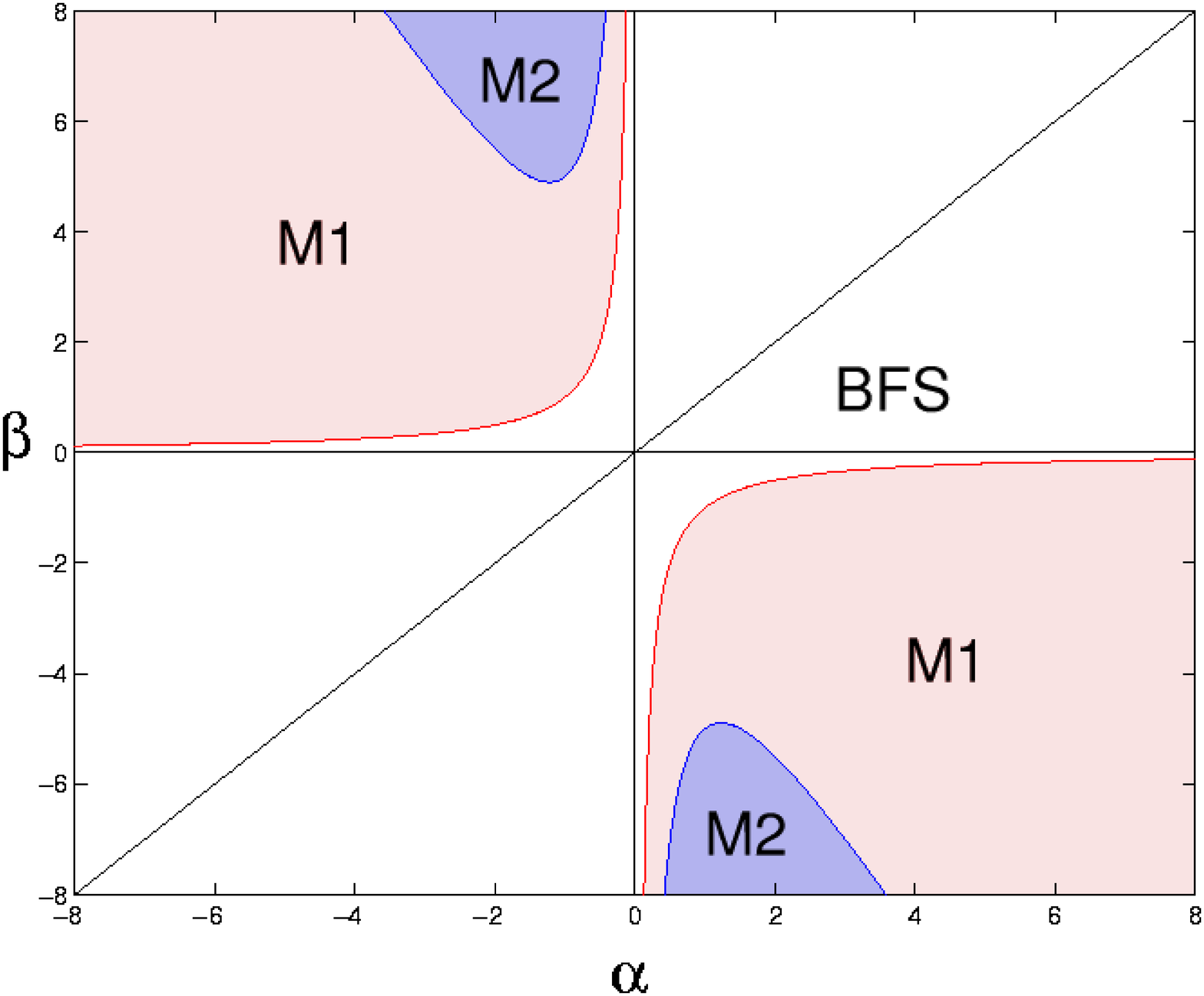} \qquad
\includegraphics[width=3.5cm]{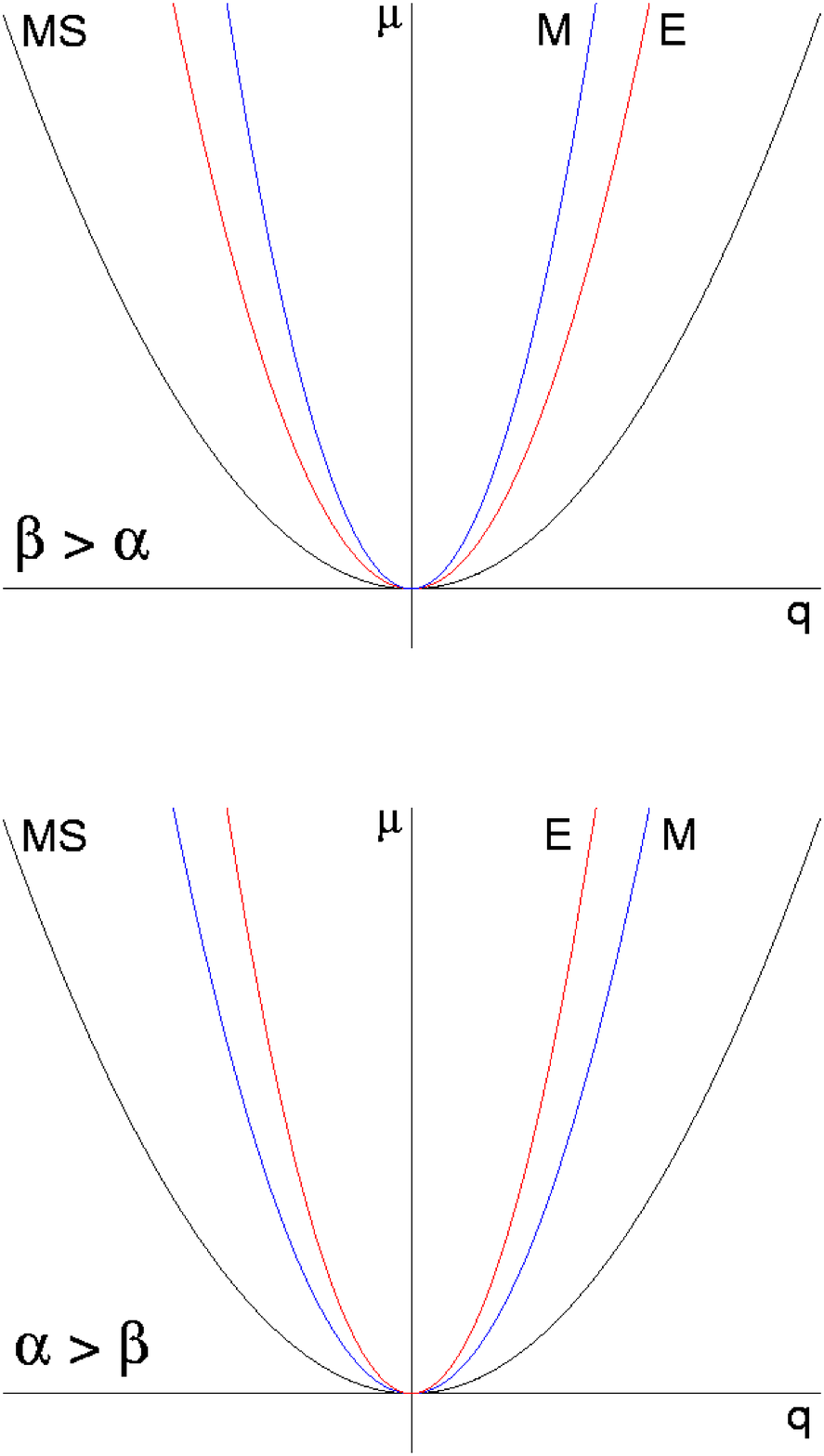} 
\end{center}
\caption{Left: wavenumber distribution
	of stationary MAWs in the $(\alpha,\beta)$ plane. 
	In (BFS), MAWs exist if $q^2>q_{\rm M}^2$.
	In (M1), MAWs exist $\forall q$.
	In (M2), MAWs exist if $q^2<q_{\rm M}^2$.
	Right: regions of existence of MAWs in the $(q,\mu)$ 
	plane in the Benjamin-Feir-Newell stable regime
	((BFS) region). (MS) is the marginal stability curve, 
	(E) is the Eckhaus instability curve and (M) is existence 
	curve defined by~(\ref{eq:defqm}). 
	Stationary MAWs exist outside (M).}
\label{fig:phases}
\end{figure}

The corresponding regions are illustrated on Fig.~\ref{fig:phases}.
Note that $q_{\rm M}(\alpha,\beta,\mu)=q_{\rm E}(\beta,\alpha,\mu)$. If
$|\alpha|=|\beta|$, the positivity of $\tilde{e}/\lambda$ is assured
when the corresponding plane wave is Eckhaus unstable. If
$|\alpha|\neq|\beta|$, the positivity does not coincide anymore with the
Eckhaus criterion; this is not surprising considering that we do not
restrict our analysis to long wavelength perturbations of plane waves,
but that the solutions we are seeking may have any wavenumber.

In the following we distinguish two cases. In the first case eigenvalue
$0$ has a simple elementary divisor, {\em i.e.}, has two distinct
eigenvectors~cite{jhale}. This coincides with case I: the MAW solution
bifurcates from the $A=0$ state, with $\omega\sim \mu \alpha$ and hence
${\tilde{e}}/{\lambda}\sim 4 \mu >0$, for $\mu>0$. In the second case,
eigenvalue $0$ has only one eigenvector. This coincides with case II:
the MAW is superimposed over a plane wave with $\omega \simeq \mu \beta$,
so $q \simeq 0$, and
\[
\frac{\tilde{e}}{\lambda} \simeq -\frac{2\mu(1+\alpha \beta)}{1+\alpha^2}>\,0,
\]
The positivity is insured if the system is Benjamin-Feir-Newell 
unstable, $(1+\alpha\beta)<0$.

In terms of $\tilde{M}, \tilde{N}, P, Q$, the {\em consistency
condition} (\ref{cc1}) can be written as
\begin{equation}
(1+\alpha^2) M = \left( \frac{a}{\lambda}\tilde{N} - \lambda Q \right)^2 
							\label{cc2}
\end{equation}
where in new variables
\begin{eqnarray*}
M & = & \frac{d(1+\alpha^2)}{2 \lambda}(d\,P+e)(\tilde{M}-d\,P^2-\tilde{e}\,P)-
\frac{d^2(1+\alpha^2)}{4}\, Q^2 \\
  &   & (\alpha \omega+\mu)(d \, P+e)^2-(1+\alpha \, \beta)(d \, P+e)^3 \,.
\end{eqnarray*}
Recalling (\ref{or2}), we may  express $G$ by
\begin{equation}
G  =  \frac{\frac{a}{\lambda}\tilde{N}-\lambda Q}{1+\alpha^2}  \,.
							\label{ag}
\end{equation}
Here we are allowed to fix the sign of the right hand side expression
because of the $(G,x)\mapsto(-G,-x)$ reflection symmetry of (\ref{or1})
and (\ref{or2}).

\subsection{Case I}

We want the eigenvalue $0$ to have non-degenerate eigenvectors, for
this, we set
\[
 \frac{\frac{\lambda}{a}}{\frac{1}{\lambda}}
=\frac{k_1}{-\frac{\tilde{e}}{\lambda}},\mbox{~i.e.,~}
  k_1=-\frac{\lambda \tilde{e}}{a}
\]
Consequently, we have 
\begin{equation}
e=\tilde{e}+\frac{a}{\lambda}k=\frac{\epsilon a}{\lambda}k_2 \,.
							\label{2nde}
\end{equation}

Notice that $e\sim 0$ to the zeroth order, so $R_0 \sim 0$ and $\omega
\sim \mu \alpha$, which means that the solution to be considered
bifurcates from the zero solution $A=0$, corresponding to a plane wave
around the marginal stability curve, with wavenumber $q \sim \pm \mu^{1/2}$. 
This solution is therefore outside the Eckhaus stability
region when $1+\alpha\beta>0$.

The four eigenvectors of $A$ are:
\[
\left(\begin{array}{c} 0\\1\\0\\0\end{array}\right) \qquad
\left(\begin{array}{c} \tilde{e}\\0\\1\\0\end{array}\right) \qquad
\left(\begin{array}{c} 0\\ -i k_1 \omega_1^{-1}\\ 1\\
      i\omega_1 \end{array}\right) \qquad
\left(\begin{array}{c}  0\\ i k_1 \omega_1^{-1}\\ 1\\
      -i\omega_1 \end{array}\right) \,.
\]
Let
\[
D=\left(\begin{array}{rccl}
        0 & \tilde{e} & 0 & 0 \\
        1 & 0 & 0 & \frac{\lambda^2}{a} \\
        0 & 1 & 1 & 0 \\
        0 & 0 & 0 & 1
        \end{array} \right),
D^{-1}=\left(\begin{array}{rccl}
        0 & 1 & 0 & -a^{-1}{\lambda^2} \\
        {\tilde{e}^{-1}}  & 0 & 0 & 0 \\
        -{\tilde{e}^{-1}} & 0 & 1 & 0 \\
        0 & 0 & 0 & 1
        \end{array} \right).
\]
and $\vec{\tilde{X}} \equiv (\tilde{x},\tilde{y},\tilde{z},\tilde{w})
= D^{-1}\vec{X}$. The dynamical equations for the new variables become
\[
\frac{d}{d\tau}\vec{\tilde{X}}=M(\omega_1)\vec{\tilde{X}}
+\epsilon\left(\begin{array}{c}
k_2(\tilde{y}+\tilde{z})+\frac{\lambda d}{a}(\tilde{y}+\tilde{z})^2\\
\frac{1}{\tilde{e}}(\tilde{x}+\frac{\lambda^2}{a}\tilde{w})(\tilde{y}+\tilde{z})\\
-\frac{1}{\tilde{e}}(\tilde{x}+\frac{\lambda^2}{a}\tilde{w})(\tilde{y}+\tilde{z})\\
-\frac{d}{\lambda}(\tilde{y}+\tilde{z})^2
\end{array}\right) \,,
\]
where
\[
M(\omega_1)=D^{-1}AD=\left(\begin{array}{rccl}
        0 & 0 & 0 & 0 \\
        0 & 0 & 0 & 0 \\
        0 & 0 & 0 & 1 \\
        0 & 0 & -\omega_1^2 & 0
                     \end{array}\right) \,.
\]

The angular frequency of the solution $\Omega$ should be close to
$\omega_1$, $\Omega^2=\omega_1^2+\epsilon \gamma$, with the shift
$\gamma$ to be determined later. Next, we change variables to:
\begin{equation}
\left\{\begin{array}{rcl}
\tilde{x} & = & x_1 \\
\tilde{y} & = & x_2 \\
\tilde{z} & = & z_1\, \sin\Omega\tau +z_2\, \cos \Omega\tau \\
\tilde{w} & = & \Omega z_1\, \cos \Omega\tau -\Omega z_2\, \sin\Omega\tau
       \end{array}\right. 
\label{eq:transfo}
\end{equation}
The 4-$D$ system of equations then takes form:
\begin{small}
\begin{eqnarray}
\dot{x_1} &=& \epsilon\,\left[ k_2 (x_2 + z_1\sin\Omega \tau + z_2\cos \Omega \tau)
		+\frac{\lambda d}{a} 
		(x_2 + z_1\sin\Omega \tau + z_2\cos \Omega \tau)^2  \right]
								\nonumber\\
\dot{x_2} &=& \frac{\epsilon}{\tilde{e}} \left[ x_1 
		+\frac{\Omega \lambda^2}{a} 
		(z_1 \cos\Omega \tau - z_2 \sin \Omega \tau) \right] 
		(x_2 + z_1 \sin \Omega \tau + z_2 \cos \Omega \tau) 
								\nonumber\\
\dot{z_1} &=& \frac{\epsilon}{\Omega} \left[ -\frac{d}{\lambda} 
		(x_2 + z_1 \sin\Omega\tau + z_2\cos\Omega\tau)^2
		\cos\Omega\tau \right.				\nonumber \\
	  & & \quad	+ \gamma (z_1 \sin\Omega\tau + z_2 \cos\Omega\tau) 
		\cos\Omega\tau 
								\nonumber \\
	  & & \quad \left. - \frac{\Omega}{\tilde{e}} \left( x_1 
		+ \frac{\Omega \lambda^2}{a} (z_1 \cos\Omega\tau 
		- z_2 \sin\Omega\tau)\right) (x_2 + z_1 \sin\Omega\tau 
		+ z_2 \cos\Omega\tau) \sin\Omega \tau \right]
								\nonumber\\
\dot{z_2} &=& \frac{\epsilon}{\Omega} \left[ \frac{d}{\lambda} 
		(x_2 + z_1 \sin\Omega\tau + z_2 \cos\Omega\tau)^2 
		\sin\Omega\tau \right.				\label{sys2} \\
	  & & \quad -\gamma (z_1 \sin\Omega\tau + z_2 \cos\Omega\tau) 
		\sin\Omega\tau 					\nonumber \\
	  & & \quad \left. - \frac{\Omega}{\tilde{e}} \left( x_1 
		+ \frac{\Omega \lambda^2}{a} (z_1 \cos\Omega\tau 
		- z_2 \sin\Omega\tau)\right)
		(x_2 + z_1 \sin\Omega\tau + z_2 \cos\Omega\tau) 
		\cos\Omega\tau \right]   
								\nonumber
\end{eqnarray}
\end{small}

The proof of the existence of weak MAWs close to $P=0$ relies
on a series of theorems from J. Hale's monograph~\cite{jhale}.
We reproduce the relevant theorems in appendix~\ref{sec:theorems},
and refer to them as the need arises.

Note that the transformation $(\tau, x_1, x_2, z_1, z_2) \rightarrow 
(-\tau,-x_1, x_2 , -z_1, z_2)$ leaves the system (\ref{sys2}) invariant. 
So, by definition \ref{def1} of appendix~\ref{sec:theorems} the system 
has the property $E$ with respect to $Q$, with 
\[
Q=\mathrm{diag}(-1,1,-1,1) \,.
\]

As we are interested only in the solutions with
definite parity, we may start the iteration with the vector
\[
\vec{X_0}=(0,a_2,0,a_4) \,.
\]
According to Theorem~\ref{sym1}, our solution 
$z(\tau,\vec{X_0},\epsilon)$ has the property
\[
Qz(-\tau,\vec{X_0},\epsilon)=z(\tau,\vec{X_0},\epsilon) \,,
\]
which means that our solutions are either symmetric or antisymmetric.
According to Theorem~\ref{sym2}, the second and the fourth determining
equations are always zero for this starting vector. For the first and
the third determining equations, the zeroth order solution of
$\vec{\tilde{X}}$ , {\em i.e.} $\vec{X_0}$, may be substituted, and we get
\begin{eqnarray}
k_2 a_2+\frac{\lambda d}{a}(a_2^2+\frac{1}{2}a_4^2) & = & 0 
							\label{nd1}\\
\frac{\gamma}{2\Omega}a_4+\frac{\lambda^2 \Omega a_2 a_4}{2a \tilde{e}}
-\frac{d} {\lambda \Omega}\, a_2 a_4 		    & = & 0  \,.
							\label{nd2} 
\end{eqnarray}
From (\ref{nd2}), we have two possibilities: either $a_4 = 0$ or
\begin{equation}
\gamma + a_2 \left( \frac{\lambda^2 \Omega^2}{a \tilde{e}}
	-\frac{2 d}{\lambda} \right) = 0 \,.		\label{nd22}
\end{equation}

When $a_4=0$, using $\vec{X_0}=(0,a_2,0,0)$ in~(\ref{sys2}) leads to a
trivial constant solution. In the following, we consider only the second
case~(\ref{nd22}). We can solve (\ref{nd1}) and (\ref{nd22}) for
$\gamma$ and $a_4$ and prove that the system (\ref{sys2}) has periodic
solutions. Note that we have three free parameters $\epsilon, a_2, k_2$.
But as we will see further, $\epsilon$ and $a_2$ are always combined as
$\epsilon a_2$ in the first approximation controlling the amplitude and
the period of the solution, and the combination will therefore be
regarded here as one single free parameter. For general periodic
solutions, $a_2$ can be interpreted as a phase control parameter, {\em
i.e.}, a parameter giving the initial location on the periodic orbit at
$\tau=\tau_0$. Here, because we only consider symmetric solutions, the
translational symmetry of the autonomous system is broken, and that is
the reason why $\epsilon$ and $a_2$ combine into a single parameter. The
remaining parameter $k_2$ can be chosen freely, for example as to
satisfy the {\em consistency condition} (\ref{cc2}), which, when the zeroth
order solution is substituted, becomes at order $(\epsilon^2)$:
\begin{equation}
-\frac{d^2}{4}\,\Omega^2\,a_4^2+\mu \left(d a_2 
	+\frac{k_2 a}{\lambda}\right)^2=0 \,.		\label{nd3}
\end{equation}

At zeroth order, $\Omega^2 = \omega_1^2 = 4\mu$ and $\tilde{e}=4\mu \lambda$. 
Solving the system of equations (\ref{nd1}),(\ref{nd22}) and (\ref{nd3}), 
we get
\[
\left\{\begin{array}{rcl} 
k_2 & =- &\frac{3\lambda}{a} d a_2 \\
\gamma & = & \frac{c}{a} a_2\\
a_4 & = & \pm 2 a_2 \end{array} \right. \,.
\]
We can write out the Jacobian for those three equations explicitly:
\[
J=\left(\begin{array}{ccc}
a_2	&  0  & \frac{\lambda d}{a} a_4 	\\ 
0	&  1  & 0 				\\
\frac{2\mu a}{\lambda}(d a_2+\frac{k_2 a}{\lambda}) 
	&  0  & -\frac{d^2}{2} \Omega^2 a_4
\end{array}\right) \,.
\]
The determinant of this Jacobian is
\[
\det J=\frac{1}{2} d^2 \Omega^2 a_2 a_4 \neq 0 \qquad a_2 \neq 0 \,.
\]

We now invoke theorem \ref{ther2}, reproduced in
appendix~\ref{sec:theorems}, and conclude our proof that system
(\ref{or1}) and (\ref{or2}) has periodic solutions near $P=0$. We shall
give approximate solutions in section~\ref{sec:approximate}, and show
that in this case they contain defects.

\subsection{Case II}

Eigenvalue 0 has only one eigenvector. In this case, we assume that
$\frac{\lambda}{a} \tilde{e}+k_1 \neq 0$ to the zeroth order in
$\epsilon$, so without loss of generality we can choose $k_2=0$. Then
$\frac{\lambda}{a}e=\frac{\lambda}{a} \tilde{e}+k_1$. Implementing the
transformation $\vec{X}=D\vec{\tilde{X}}$ with
\[
D=\left(\begin{array}{cccc}
        0 & \tilde{e} & 0 & 0 \\
        1 & 0 & 0 & - k_1\omega_1^{-2} \\
        0 & 1 & 1 & 0 \\
        0 & 0 & 0 & 1
        \end{array} \right)
\]
we have
\[
\frac{d}{d\tau}\vec{\tilde{X}} = 
	M(\omega_1)\vec{\tilde{X}}+\epsilon\left(\begin{array}{c}
	-\frac{d}{\tilde{e}} k_1 (\tilde{y}+\tilde{z})^2 \\
\frac{1}{\tilde{e}}(\tilde{x}-k_1
	\frac{\lambda}{\tilde{e}}\tilde{w})(\tilde{y}+\tilde{z}) \\
-\frac{1}{\tilde{e}}(\tilde{x}-k_1
	\frac{\lambda}{\tilde{e}}\tilde{w})(\tilde{y}+\tilde{z}) \\
-\frac{d}{\lambda}(\tilde{y}+\tilde{z})^2
\end{array} \right) \,,
\]
where
\[
M(\omega_1)=D^{-1}AD=\left(\begin{array}{rccl}
        0 & \frac{\lambda e}{a} & 0 & 0 \\
        0 & 0 & 0 & 0 \\
        0 & 0 & 0 & 1 \\
        0 & 0 & -\omega_1^2 & 0
                     \end{array}\right) \,.
\]

As in case I, let $\Omega^2=\omega_1^2+\epsilon \gamma$ and perform the
same transformation~(\ref{eq:transfo}) into variables $x_1,x_2,z_1,z_2$.
We then obtain a 4-$D$ system similar to (\ref{sys2}). However, in the
equation for $\dot{x_1}$, there is an $\epsilon$-free term. In order to
use the successive approximation method, further transformations are
required. Let $\rho \in \mathbb{R}$ such that $\rho^2=\epsilon$. With
the transformation $x_2 \rightarrow \rho x_2, \epsilon \rightarrow
\rho^2$ we recover the standard form

\begin{small}
\begin{eqnarray}
\dot{x_1} & = & \frac{\rho \lambda e}{a} x_2-\rho^2 k_1\frac{d}{\tilde{e}}
		(\rho x_2+z_1\, \sin\Omega \tau+z_2 \,\cos \Omega \tau)^2 
								\nonumber\\
\dot{x_2} & = &\frac{\rho}{\tilde{e}}
		\left[ x_1-k_1 \frac{\Omega \lambda}{\tilde{e}} 
		(z_1\, \cos\Omega\tau - z_2 \sin\Omega\tau) \right]
		(\rho x_2+z_1 \sin\Omega\tau + z_2 \cos\Omega\tau) 
								\nonumber\\
\dot{z_1} & = & \frac{\rho^2}{\Omega} 
		\left[ \frac{-d}{\lambda} 
		(\rho x_2 + z_1 \sin\Omega\tau + z_2\cos\Omega\tau)^2
		\cos\Omega\tau \right. \\			\nonumber \\
	  &   & + \gamma (z_1\sin \Omega \tau+z_2 \cos \Omega \tau) \cos\Omega\tau
								\label{sys33} \\
	  &   & - \frac{\Omega}{\tilde{e}}  \left.
		\left(x_1-\frac{\Omega \lambda}{\tilde{e}} 
		k_1\,(z_1\, \cos\Omega\tau - z_2 \sin\Omega\tau)\right)
		(\rho x_2+z_1 \sin\Omega\tau + z_2 \cos\Omega\tau) 
		\sin\Omega\tau \right]
			 					\nonumber \\
\dot{z_2} & = & \frac{\rho^2}{\Omega} \left[ \frac{d}{\lambda} 
		(\rho x_2+z_1 \sin\Omega\tau + z_2 \cos\Omega\tau )^2 
		\sin\Omega\tau \right.
								\nonumber \\
	  &   & - \gamma (z_1\sin\Omega\tau + z_2\cos\Omega\tau) \sin\Omega\tau 
								\nonumber \\
	  &   & - \frac{\Omega}{\tilde{e}} \left.
		\left( x_1 -\frac{\Omega \lambda}{\tilde{e}} k_1 
		(z_1\, \cos\Omega\tau -z_2\sin\Omega\tau) \right) 
		(\rho x_2 + z_1 \sin\Omega\tau + z_2\cos\Omega\tau) \cos\Omega\tau 
		\right] \,,
								\nonumber\\
\end{eqnarray}
\end{small}
The system (\ref{sys33}) has the same symmetry as identified in the case
I. If we are only interested in solutions with definite parity, we may
again start the iteration with $\vec{X_0}=(0,a_2,0,a_4)$. To the second
order $(\rho^2)$, the determining equations are:
\begin{eqnarray}
a_2 \frac{\lambda e}{a}-\rho \frac{d a_4^2 k_1}{2\tilde{e}}	
		+0(\rho^3)&=&0			\label{dd1} \\
\rho \frac{\gamma a_4}{2 \Omega}-\rho^2\left(\frac{d a_2 a_4}{\lambda \Omega}
		+\frac{\lambda \Omega a_2 a_4 k_1} {2\tilde{e}^2}\right)
		+0(\rho^3)&=&0 \,.		\label{dd2}
\end{eqnarray}
From the second equation we obtain either $a_4=0$ (trivial for
our purposes, as discussed above) or
\begin{equation}
\gamma-\rho a_2\left(\frac{2 d}{\lambda}
+\frac{\lambda \Omega^2 k_1}{\tilde{e}^2}\right)+0(\rho^2)=0 \,.
						\label{dd22}
\end{equation}
If we backtrack the transformations made, it is clear that 
the {\em consistency condition} requires that we keep terms up to the 
fourth order $(\rho^4)$. We found that with the substitution
\[
e=\frac{\alpha \omega + \mu}{1+\alpha \beta}
+\rho^2(\rho^2 \omega_3-\rho d a_2) \,,
\]
where $\omega_3$ is a new parameter, only the fourth or higher order
terms are left in the {\em consistency condition}. From the definition $e =
R_0^2 = (\omega-\mu \alpha)/(\beta-\alpha)$ and the above equation, we
get $\omega \sim \mu \beta$ and then $e \sim \mu$ to the zeroth order.
So $R_0 \sim \sqrt{\mu}$, $q \sim 0$, which means that this solution
bifurcates from the HOS $A=\sqrt{\mu}\exp({-i \omega t})$. To the leading
order ($\rho^4$), we are allowed to use the following substitutions in
the {\em consistency condition} (\ref{cc2}):
\begin{eqnarray}
a_2 \rightarrow 0 \quad 
\omega \rightarrow \mu \beta \quad 
\Omega \rightarrow \sqrt{-\frac{2\mu(1+\alpha \beta)}{1+\alpha^2}} ,
							\nonumber \\ 
k_1 \rightarrow \frac{\mu \lambda}{a}
\left( 1+\frac{2 \lambda (1+\alpha \beta)}{1+\alpha^2}\right) \quad
\tilde{e} \rightarrow -\frac{2 \mu \lambda (1+\alpha \beta)} {1+\alpha^2} \,.
							\label{app}
\end{eqnarray}
The resulting equation is of a relatively simple form:
\begin{equation}
a_4^2(-\lambda+d^2(1+\alpha \beta)(1+\alpha^2+\lambda
+\lambda \alpha \beta))+4(1+\alpha \beta)^2 \lambda \mu \omega_3 = 0 \,.
							\label{dd3}
\end{equation}
From (\ref{dd1}) it follows that $a_2$ is of order $\rho$, and from 
(\ref{dd22}) that $\gamma \sim 0(\rho^2)$. After a change of 
variable $a_2=\rho \, a_{22}$ and keeping only the highest 
order for the equations, we can rewrite (\ref{dd1}) and (\ref{dd22})
as
\begin{eqnarray}
a_{22} \frac{\lambda e}{a}-\frac{k_1 d a_4^2}{2 \tilde{e}} & = & 0  \label{ndd1} \\
\gamma 							   & = & 0 .\label{ndd2}
\end{eqnarray}
For $e, \tilde{e}, k_1$ we use the values in (\ref{app}). 
From (\ref{ndd1}), (\ref{ndd2}) and (\ref{dd3}), we can solve for
$a_{22}, \gamma, \omega_3$. The Jacobian of those equations is
\[
J=\left(\begin{array}{ccc}
    \frac{\lambda e}{a} & 0  & 0 \\
       0                & 1  & 0 \\ 
       0 & 0 & 4(1+\alpha \beta)^2 \lambda \mu
   \end{array} \right), 
\]
So, $\det J = 4(1+\alpha \beta)^2 \lambda^2 \mu e / a \neq 0$ 
for $1+\alpha \beta \neq 0$. According to Theorem~\ref{ther2}, 
we have proved that equations (\ref{or1}) and (\ref{or2})
possess periodic solutions.

\section{Analytic form of periodic solutions, stability analysis
	and numerical tests}
\label{sec:approximate}

We have proved in the preceding section the existence of symmetric
periodic solutions in case I and II. In both cases, a small parameter
$\epsilon$ or $\rho$ ensures the convergence of successive
approximations. However, we did not give a bound on the highest value of
this parameter, nor did we show that the solutions which we obtain are
the ones observed in numerical simulations. In this section we give the
approximate analytical form of periodic solutions. We 
compare them with direct numerical integration of the CGLe
in case II. 

The solutions are shown to be independent of $\lambda$ 
to order $\epsilon$ in case I
and to order $\epsilon^2$ in case II. In addition, these solutions
should also satisfy the 3-$D$ ODE mentioned in
section~\ref{sec:intro:MAWs} which do not contain $\lambda$, so they can
be matched with the solutions of the 3-$D$ system in a unique way,
independent of the value of $\lambda$. Hence, we conclude that to all
orders the physical solutions are identical for the two values of
$\lambda$.

The two cases are taken separately. In this section, we reinstate
$x$ as the spatial variable, $R=R(x)$.

\subsection{Case I}

Using (\ref{dR}), (\ref{ag}) and the case I calculations of the
preceding section, we have after some algebra:
\begin{eqnarray}
R^2 & = & -2 \epsilon d a_2(1\pm \cos \Omega x) 
						\label{rphi} \\
\phi_x & = & -\frac{\epsilon a_2}{2(1+\alpha^2)\Omega}
\frac{\sin2\Omega x \pm 2\sin\Omega x}{1\pm \cos \Omega x} \,. 
						\nonumber
\end{eqnarray}
To the first order of $\epsilon$, $R$ and $\phi_x$ 
are independent of $\lambda$. The
$\pm$ sign selects two solutions which transform into each other by
translating by a half period. This is reminiscent of the spatial
translational invariance in the symmetric solution space. From the
definitions of $e, \Omega$ and from (\ref{2nde}), (\ref{domega1}), we
get to the first order:
\begin{eqnarray}
\omega &=& \mu \alpha-3 \epsilon  a_2 \nonumber\\
\omega_1^2 &=& 4 \mu+\frac{6 \epsilon d a_2}{1+\alpha^2}(\alpha \beta +2 \alpha^2+3)
	\nonumber\\
\Omega &=& \omega_1+\frac{\epsilon \gamma}{2 \omega_1} \,.
							\label{sln1}
\end{eqnarray}

We see that $\omega$ and $\Omega$ are independent of $\lambda$. On the
other hand, for periodic boundary conditions, we can use Fourier modes
directly to transform the PDE (\ref{cgl}) to a finite set of approximate
ODE's by Galerkin truncation. Then the stationary solution can be
obtained by solving a set of nonlinear algebraic equations.

\subsubsection*{Numerical comparison}

If we take as an example the following parameter values (previously
used in~\cite{Lega:holes}) for which defect chaos is expected:
\[
\alpha=1.5,\quad  \beta=-1.2
\]
and fix the size of the domain to $L=24$, then at $\mu=0.072644$,
$\omega=0.097879$, a periodic solution of period $L/2$ is found. This
solution has $R_{max} \simeq 0.0750$. On the other hand, if we use the
same $\alpha,\beta,\mu$ and search for $R_{max} \simeq 0.075$ by
adjusting $\epsilon$ (we always keep $a_2=1$), we find that 
\[
\epsilon \sim 0.00380,\quad \omega=0.097566, 
\quad \mbox{ period } \frac{2\pi}{\Omega}=12.0102 \,.
\]

\begin{figure}[ht] 
\begin{center}  \includegraphics{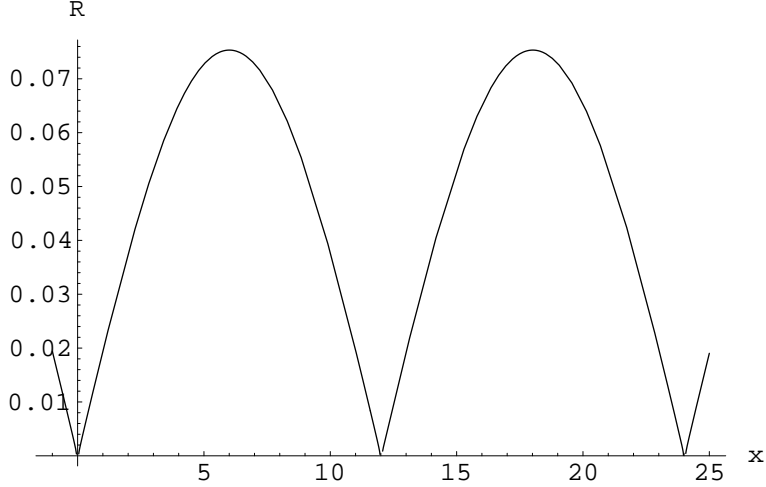} \end{center}
\caption{Spatial profile of the amplitude $R(x)$ at $\mu=0.072644$,
	from (\ref{rphi}) with $R_{max} = 0.075$. 
\label{fig1}} \end{figure}

The approximate analytic solution and the numerical solution of the
exact CGLe agree very well. The profile of $R$ from our successive
approximation is shown in Fig.~\ref{fig1}.

\subsubsection*{Structure near the defect}

It is easy to see from (\ref{rphi}) that only $\epsilon a_2>0$ is the
physically interesting combination. However, we may wonder whether it is
really true that $R^2=dP+e$ remains non-negative everywhere while
touching zero at some points. Fig.~\ref{fig1} and the first equation of
(\ref{rphi}) suggest a positive answer to this question. But since we
have only an approximate solution, further justification is needed.
Suppose at some instant $x_0$, we have $dP+e=0$ on the periodic orbit.
From the {\em consistency condition} (\ref{cc2}), at this transition point
\[
\frac{d^2(1+\alpha^2)}{4}Q^2 
+ \left(\frac{a}{\lambda}\tilde{N}-\lambda Q\right)^2=0,
\]
so, $Q=\tilde{N}=0$. According to (\ref{eq:4d}), $\dot{\tilde{M}}=0$ and
$\dot{P}=0$. Assume that $\dot{Q}=0$, then $\dot{\tilde{N}}\neq 0$ since
the point is not an equilibrium. At next instant $x_0+\delta x$, the
{\em consistency condition} can not be satisfied as the two sides of
(\ref{cc2}) have different orders of $\delta x$. So we conclude that
$\dot{Q}\neq 0$ at the point $x_0$, which means that $Q(x_0+\delta x)$
has negative sign to that of $Q(x_0-\delta x)$. Thus, after touching the
zero value plane, $dP+e$ returns to the positive half space again. The
turning happens exactly on the $dP+e=0$ plane. We claim that $dP+e \geq
0$ always holds and the equality holds periodically. From (\ref{rphi}),
in the neighborhood of $R=0$ at $x=x_0$ on the periodic orbit, $R$
behaves like 
\[
R \sim \left(\frac{d \dot{Q}}{2} \right)^{1/2} |x-x_0| \,,
\]
and is manifestly a non-analytic function of $x$.

We do not discuss the stability of the solutions in case I, as
this has already been accomplished by Tak\'{a}\v{c}~\cite{pt} who has
proven that these solutions are unstable.

\subsection{Case II}

To the first order of $\epsilon$, the solutions are
\[ \left \{ \begin{array}{rcl}
    x_1 &=&  -\frac{\epsilon k_1 a_4^2}{8\tilde{e}^2 \Omega}(2d\tilde{e}
		+\lambda k_3)\sin2 \Omega x \\
    x_2 &=& \epsilon (a_2^2-\frac{\lambda k_1 a_4^2}{4\tilde{e}^2}\cos2 \Omega x) \\
    z_1 &=& -\frac{\epsilon a_4^2}{12 \tilde{e}^2 \lambda \Omega^2}
		(3(3d\tilde{e}^2+\lambda^2 \Omega^2 k_1)\sin \Omega x
		+(d\tilde{e}^2-\lambda^2 \Omega^2 k_1)\sin 3\Omega x) \\
    z_2 &=& a_4+\frac{\epsilon a_4^2}{12\tilde{e}^2\lambda \Omega^2}
		(3(\lambda^2 \Omega^2 k_1-d\tilde{e}^2)\cos\Omega x
		+(\lambda^2 \Omega^2 k_1-d\tilde{e}^2)\cos 3\Omega x) \,, \\
   \end{array}\right.
\]
where $\epsilon=\rho^2>0$, and $a_4$ is a free parameter. In the
following, we will see that $\epsilon$ and $a_4$ always emerge 
in the combination $\epsilon \, a_4$. To the second order, 
$\omega$ is
\[
\omega=\mu \beta +\frac{\epsilon^2 a_4^2}{4\mu(1+\alpha^2)}
	\left(\frac{1+\alpha \,\beta}{\beta-\alpha} 
	+\frac{\beta-\alpha}{1+\alpha \,\beta}\right) \,.
\] 
It is independent of $\lambda$, and therefore $e,b,\Omega$ are also
independent of $\lambda$. $R$ and $\phi_x$ can also be calculated 
to the second order:
\begin{eqnarray}
R^2 &=& -\frac{d^2}{2\mu}(\epsilon a_4)^2 + d \epsilon a_4 \cos\Omega x
	+\frac{d(\epsilon a_4)^2}{12 \Omega^2} 
	\left( \frac{c}{a}+\frac{e}{b}\right) \cos2\Omega x + e 
						\label{sol:case2} \\
\phi_x &=& \frac{\epsilon a_4}{\mu \Omega (1+\alpha^2)}
	\left[e \sin\Omega x-\frac{\epsilon a_4}{24\Omega^2}
	\left(6d\Omega^2+\frac{7e^2}{a\Omega^2}+\frac{7ce}{a}\right)
	\sin2 \Omega x\right]
						\nonumber 
\end{eqnarray}

So clearly $R$ and $\phi_x$ are independent of $\lambda$. Similarly, the
different signs of $a_4$ will give the same solution up to a half-period
translation. This solution is the one observed in the numerics when
passing the Eckhaus instability for underlying wavevector $q=0$. Linear
stability analysis reveals~\cite{stp} that the $q=0$ state, the most
stable state under the long wavelength perturbations, becomes unstable
when the size of the system is such that the smallest possible nonzero
wavenumber $k$ satisfies 
\[
k^2 < -\frac{2\mu(1+\alpha \beta)}{1+\alpha^2}\equiv \kappa^2 \,.
\]
It is easy to see that $\kappa^2=\omega_1^2$ up to order $(\rho^4)$. 

For our parameter choices $\mu=1, \alpha=1.5, \beta=-1.2$,
the bifurcation size of the system is $L_0=\frac{2\pi}{\kappa}=8.95492$.
In the following, we will first prove the stability of our solutions 
near the bifurcation point. Then we will compare them with the stable
solutions observed in numerics.

\subsubsection*{Stability analysis: presentation}

Assume that $A=R\exp({i\phi})$ where $R, \phi \in \mathbb{R}$ is 
an exact solution of (\ref{cgl}). The perturbed solution is assumed to be 
$\bar{A}=(R+r)\exp({\phi+\theta})$, where $r, \theta \in
\mathbb{R}$ is the perturbation on the amplitude and phase, separately. 
Substitute it into (\ref{cgl}), keeping only the linear terms in $r$ 
and $\theta$. We have 
\begin{eqnarray}
r_t & = & (\mu-\phi_x^2-\alpha \phi_{xx}-3 R^2)r
		+ r_{xx}-2 \alpha \phi_x r_x \nonumber \\
    &   & -(2 R \phi_x+2 \alpha R_x)\theta_x-\alpha R \theta_{xx} 
						\label{st1} \\
R \theta_t & = & (\omega-\alpha \phi_x^2+\phi_{xx}
		-3 \beta R^2)r+\alpha r_{xx}+2 \phi_x r_x \nonumber \\
    &   & + (2 R_x-2 \alpha R \phi_x)\theta_x+R \theta_{xx}  \,,
						\label{st2} 
\end{eqnarray}
where in (\ref{st2}) we have used $\phi_t=-\omega$. To study the
stability of the starting solution $A$, we treat these equations as an
eigenvalue problem for a two components vector, {\em i.e.}, we let
$r_t=\sigma r$, $\theta_t=\sigma \theta$ and we investigate the spectra
$\sigma$ of the linear operator resulting from (\ref{st1}) and
(\ref{st2}) in the $C^1$ continuous periodic function space. As the CGLe
has global phase invariance, the eigenvalue equations always have
solution $(r,\theta)=(0,\theta_0)$ with eigenvalue
$\sigma=0$. At the same time, spatial translational invariance implies
that another eigenmode has $\sigma=0$. As a result, saying that the
solution is stable means that it is stable up to a phase and a spatial
translation, and that all other eigenmodes have eigenvalues with
negative real parts.

Invoking the expression for $R,\phi_x$ to the second order of
$\epsilon$, the coefficients of various terms of $r,\theta$ and their
derivatives in (\ref{st1}) and (\ref{st2}) become explicit functions of
$x$. The resulting linear operator on $(r,\theta)$ has even parity due
to the symmetry of our solution, and we can consider the even and odd
solutions of $r,\theta$ separately. If we set $\epsilon=0$, i.e., the
starting state $A$ is a plane wave state, then $\cos(n\Omega x)$ and
$\sin(n\Omega x)$ are the eigenfunctions of the unperturbed linear
operator. They give the stability spectrum of the plane waves. Now, let
us move a little (to the order of $\epsilon$) beyond the bifurcation
point. The eigenfunctions are still $\cos(n\Omega x)$ and $\sin(n\Omega
x)$ up to $\epsilon$ corrections. For example, if the even solutions are
considered first, we assume that to the first order the eigenfunctions
are (the time dependence for $r,\theta$ has been suppressed):
\begin{eqnarray}
r & = & m_1\cos(n\Omega x)+\epsilon (m_0\cos((n-1)\Omega x)
	+ m_2\cos((n+1)\Omega x)) 	\label{eig1} \\
\theta & = & n_1\cos(n\Omega x)+\epsilon (n_0\cos((n-1)\Omega x))
	+ n_2\cos((n+1)\Omega x)) \,,	\label{eig2}
\end{eqnarray}
where $n$ is a non-negative integer. Note that we do not include the
terms such as $\epsilon^2 \cos((n\pm 2)\Omega x)$ in the above
expressions because they induce corrections of order $\epsilon^3$ or
higher in the eigenvalues. Now if we substitute (\ref{eig1}) and
(\ref{eig2}) into the eigenvalue equations and identify the coefficients
of $\cos(n\Omega x), \cos((n-1)\Omega x) \mbox{ and } \cos((n+1)\Omega
x)$, a set of six homogeneous linear equations for
$m_0,m_1,m_2,n_0,n_1,n_2$ can be derived. The determinant of the
coefficient matrix will give an eigenvalue equation for $\sigma$. The
resulting expression is too complicated to merit being displayed here.

Before bifurcation, the HOS is stable. The first instability occurs for
$n=1$ mode, one eigenvalue of which is very close to $0$ near the
bifurcation point, being negative before and positive after. Meanwhile,
for $n>1$ modes, the corresponding eigenvalues have negative real parts
bounded away from zero. As the bifurcating solution emerges continuously
from the HOS, near the bifurcation point $(\epsilon \ll 1)$ the
perturbed linear operator has all the eigenvalues with negative real
parts away from $0$ for $n>1$ and one eigenvalue close to $0$ for $n=1$.
So, we only need to check the stability of our solutions for $n=1$. 

For convenience, we can fix parameters $\alpha$ and $\beta$ to any
values allowed by~(\ref{eq:positivity}) and perform the above 
stability analysis of the solution.

\subsubsection*{Stability analysis: numerical checks}

The numerical values we used are 
$\mu=1.0, \alpha=1.5,\beta=-1.2,a_4=1$. The eigenvalue equation is then
\begin{eqnarray*}
            7.9860 \epsilon^2 \sigma 
+ (56.423 - 63.394 \epsilon^2)\sigma^2 
+ (82.564 - 75.135 \epsilon^2)\sigma^3 & & \\
+ (45.022 - 28.859 \epsilon^2)\sigma^4 
+ (10.923 - 4.3059 \epsilon^2)\sigma^5 & & \\
+ (1.0    - 0.18864\epsilon^2)\sigma^6 &=& 0 \,.
\end{eqnarray*}
$\sigma=0$ corresponds to the neutral mode associated with the global
phase invariance. All others solutions have negative real parts. The
$\sigma_{-}=-0.14154 \epsilon^2$ solution is the interesting one. If we
use the same parameter values to calculate the stability of the HOS, the
eigenvalue equation for $n=1$ is
\[
\epsilon^2(-0.21122 - 0.26402\sigma) + 2.98462 \sigma + \sigma^2=0 \,.
\] 

To the second order in $\epsilon$, we have $\sigma=-2.98462- 0.19325
\epsilon^2$ or $\sigma_{+}=0.07077\epsilon^2$. The later positive
eigenvalue indicates that the plane wave solution is not stable. We note
that $2\sigma_{+}=-\sigma_{-}$ to order $\epsilon^2$ which
indicates a supercritical pitchfork bifurcation. We have proved
that this equality holds exactly at the bifurcation point
for any values of $\alpha$ and $\beta$, and this justifies the 
above numerical checks.
Under perturbation the HOS will evolve to the
modulated amplitude solution given above. When the instability is
saturated, the corresponding eigenvalue for the MAW is negative. If we
change the sign of $a_4$ or use the other value of $\lambda$, the
eigenvalue does not change, as expected. 

If we alternatively consider
the odd-parity function space $\{\sin(n\Omega x)\}_{n \in \mathbb{N}}$,
we obtain the following eigenvalue equation:
\begin{eqnarray*}
 (28.2115-33.6568 \epsilon^2)\sigma + (27.1763-21.2703\epsilon^2)\sigma^2 
+ (8.92308 - \\
 4.04125\epsilon^2)\sigma^3 + (1 -0.19287\epsilon^2)\sigma^4 =0 \,.
\end{eqnarray*}
This equation is quartic because for $n=1$ only two modes $\sin \Omega
x$ and $\sin 2\Omega x$ are used. Now $\sigma=0$ corresponds to the
neutral mode associated with the spatial translation of the CGLe. Other
eigenvalues of the equation have negative real parts bounded away 
from $0$. 

To summarize, our solution is stable in the whole phase space of 
the CGLe, up to a phase and a spatial translation.

In ref~\cite{sup}, B. Janiaud {\it et al.} have investigated the
stability of traveling waves near the Eckhaus instability in
Benjamin-Feir stable regime. They derived a necessary condition for the
bifurcation to be supercritical and located the corresponding regions as
two strips in the $\alpha, \beta$ parameter space. We have studied the
stationary MAWs in the Benjamin-Feir unstable regime and found that the
bifurcation from the HOS to MAWs is always supercritical, even when
parameter values lay outside of the region given in ref~\cite{sup}.

In ref~\cite{quarter2}, application of the perturbation method to the
zeroth order $(\epsilon^0)$ equation gave nonzero eigenvalue
$\lambda_0=2/\beta$. This can not be correct since the zeroth order
equation just gives the stability of the unstable HOS. Furthermore, in
the Galerkin projection, somewhat surprisingly the $N=1$ truncation was
found to give a better result than the $N=2$ truncation. In our case, if
we use only the first order expressions for $R,\phi_x$ in (\ref{st1})
and (\ref{st2}), we cannot get the correct eigenvalues even near the
bifurcation point, not to mention that it would not be possible to
extend the result to the next bifurcation.

\subsubsection*{Comparaison with numerical integration of the CGLe}

\begin{figure}[ht] 
\begin{center} \includegraphics{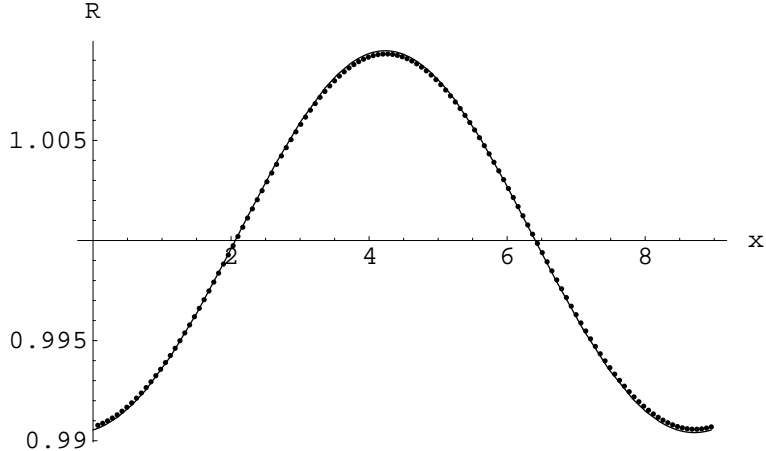} \end{center} 
\caption{Spatial profiles of the amplitude $R$ for $\mu=1, \alpha=1.5, 
	\beta=-1.2, L=8.958$ from numerical simulation (dots) and 
	the approximate solution~(\ref{sol:case2}) (solid line). The 
	agreement is good, with the discrepancy mainly due to the long 
	relaxation time close to the bifurcation. 
\label{fig2}} 
\end{figure}

In our numerical simulations we employed a pseudo-spectral method to
evolve equation (\ref{cgl}) using $128$ modes. For system size $L<L_0$,
we always recover the HOS ($q=0$). For $L$ slightly larger than $L_0$,
however, the solution relaxes to the modulated amplitude solution given
irrespective of the initial condition. Figure~\ref{fig2} depicts the
stable steady solutions given by the two methods.

\section{Conclusion}

We have reformulated the stationary one-dimensional CGLe in a finite box
with periodic boundary conditions as a fourth-order ODE for a variable
$P$ that can be interpreted as the modulation of the amplitude squared
of a plane wave solution. This reformulation enabled us to prove the
existence of stationary MAW solutions in the two limit cases
corresponding to the bifurcation of the trivial solution $A=0$ (case I),
or to the bifurcation of the plane wave solution of zero wavenumber
(case II), when those solutions are within the Benjamin-Feir-Newell
regime, or more generally in a region of instability defined
by~(\ref{eq:positivity}). That region coincides with the Eckhaus domain
if $|\alpha|=|\beta|$, but it is different otherwise. We proved the
stability of MAW solutions for the full CGLe in a finite box with
periodic boundary conditions in case II, where a homogeneous plane wave
becomes unstable. We tested our analytical results by comparison of
numerical integrations of the full CGLe with our approximate analytical
solutions.

In case I, unstable periodic hole solutions were shown to exist. This
could not be inferred from any phase equation: around the defect point
$A=0$, the amplitude behaves non-analytically, namely piecewise
affinely, and the phase is not defined. In case II we found the
symmetric stable solutions observed in the numerical integrations of the
CGLe just beyond the bifurcation point, using the box size $L$ as the
bifurcation parameter. This bifurcation was shown to be always
supercritical in the Benjamin-Feir-Newell unstable regime. The MAWs
continue to exist when the size $L$ is increased.

The analysis of MAWs bifurcating from a plane wave with wavenumber
$0<q<1$ should be similar to the study of case II. It would be
interesting to study the higher order instabilities of MAWs when the
system size is increased beyond the region in which our analysis takes
place. It has been observed that stationary symmetrical MAWs bifurcate
into uniformly-propagating asymmetrical ones via a drift-pitchfork
bifurcation. This happens when $L$ is increased as a consequence of the
growth of the amplitude of the modulation, and the increase of the
spectral richness of the MAW solution. Moreover, MAWs are expected to be
the building blocks of phase turbulence, and the analytical analysis of
their global stability may lead to a characterization of the suspected
transition between phase and defect chaos in the CGLe~\cite{maw,defect}.

{\bf Acknowledgments} The authors thank Georgia Tech. Center
for Nonlinear Science and G.P. Robinson for support. Conversations
with J. Lega are gratefully acknowledged.

\appendix
\section{Derivation of the governing equation}
\label{sec:algebra}

We use (\ref{dC}) and (\ref{dR}) to rewrite (\ref{rd}) using only $P$ and 
its spatial derivatives:
\begin{equation}
	P(aP_{xxx}+bP_x+cPP_x)_x-P_x(aP_{xxx}+bP_x+cPP_x) = (dP+e)P^3  \label{eq:*}
\end{equation}
Note that this equation contains even numbers of derivatives of $P$ in each
term in parenthesis, and also that the powers of $P$ increase while the
derivatives decrease. We now rewrite the equation in a form which take
advantage of this structure. For example, the following equation is
equivalent to (\ref{eq:*}) for any real $\lambda$:
\begin{eqnarray*}
	P(aP_{xxx}+bP_x+(c+\lambda)PP_x)_x-P_x(aP_{xxx}+bP_x+(c+\lambda)PP_x)\\
	=P^2(\lambda P_{xx}+dP^2+eP) \,,
\end{eqnarray*}  
or, put in another form and introducing another real parameter $k$:
\begin{eqnarray*}
\left(\frac{(aP_{xx}+bP+\frac{c+\lambda}{2}P^2 )_x}{P}\right)_x
=\lambda P_{xx}+dP^2+\tilde{e}P+\frac{a}{\lambda}kP \,,
\end{eqnarray*} 
where we have written $\tilde{e}+\frac{a}{\lambda}k=e$.

In this equation, we have three free parameters: besides $\omega$,
introduced by the ansatz~(\ref{eq:antsatz}) as the carrier frequency 
of the solution, we have introduced free parameters $\lambda$ and $k$. 
We now fix $\lambda$ by imposing the condition
\begin{eqnarray}
\frac{a}{\lambda}=\frac{b}{\tilde{e}}=\frac{c+\lambda}{2d} \label{condi} \, ,
\end{eqnarray} 
which allows us to write the equation in a more suggestive form:
\begin{equation*}
\left(\frac{(\lambda P_{xx}+dP^2+\tilde{e}P )_x}{P}\right)_x
=\frac{\lambda}{a}(\lambda P_{xx}+dP^2+\tilde{e}P)+kP \,, 
\end{equation*} 
the equation (\ref{ode1}) that leads to the 4-$D$ ODE of
section~\ref{sec:existence}.

$\lambda$ is determined by (\ref{condi}):
\begin{equation}
\lambda^2+c\lambda-2ad=0 \label{l1} \,.
\end{equation}
The discriminant of (\ref{l1}) is
\begin{eqnarray*}
\Delta & = & c^2+8ad \\
       & = & \left(\frac{1+\alpha^2}{2(\beta-\alpha)}\right)^2\left(\frac{9(1+\alpha \beta)^2}
{(\beta-\alpha)^2}+8\right) \,.
\end{eqnarray*}
So $\Delta>0$ for any real values of $\alpha$ and $\beta$, and the quadratic equation
(\ref{l1}) always has two real roots
\begin{eqnarray}
\lambda=\frac{3(1+\alpha \beta)(1+\alpha^2)}{4\,(\beta-\alpha)^2} \pm \frac{1+\alpha^2}
{4(\beta-\alpha)^2} \sqrt{9(1+\alpha \beta)^2+8 (\beta-\alpha)^2} \label{lam}
\end{eqnarray}

Note that $\lambda$ is a function of $\alpha$ and $\beta$ only. In some
applications~\cite{per}, the two values of $\lambda$ correspond to two
distinct solutions of the CGLe. In our case, $\lambda$ is an
intermediate variable used in the derivation and the proofs, but our
solutions to the CGLe do not distinguish the two values of $\lambda$.

\section{Theorems used in the proofs}
\label{sec:theorems}

We use successive approximation method to prove the existence of
modulated amplitude waves. Below are listed several theorems from the
theory of nonlinear oscillations taken from Hale's
monograph~\cite{jhale}. 

Consider the system of equations
\begin{equation}
\dot{z}=Az+\epsilon Z(\tau,z,\epsilon) \label{sys1}
\end{equation}
where $A$ is a constant matrix, $\epsilon,\tau \in \mathbb{R}$, and $z,Z
\in \mathbb{R}^n$. $Z$ is a continuous function of $\tau,z,\epsilon$,
periodic in $\tau$ of period $T$. In the following, we only consider the
case that $Z$ is a smooth function. Without loss of generality, $A$ can
always be assumed to have the standard form
\[
A=\mathrm{diag}(0_p,B),
\]
Where $0_p$ is a $p\times p$ zero matrix and $B$ is a constant matrix
with the property that the equation $\dot{y}=By$ has no nontrivial
periodic solution of period $T$. Under these settings, if the successive
approximation is applied to (\ref{sys1}), we have
\begin{theorem} \label{ther1}
Given $d>b>0$, there is an $\epsilon_1>0$ such that for any given 
constant $p$ vector $a,\| a \| <b$ and real $\epsilon, 
|\epsilon|<\epsilon_1$, there is a unique function
\[
z^*(\tau)=z(\tau,a,\epsilon), \mbox{with } \underset{\tau}{\sup}\|z^*(\tau)\|<d
\] 
which has continuous first derivative with respect to $\tau$ and satisfies
\[
\dot{z^*}=Az^*+\epsilon Z(\tau,z^*,\epsilon)-\epsilon P_0 Z(\tau,z^*,\epsilon).
\]
Furthermore, $z(\tau,a,0)=a^*$, $a^*=\mathrm{col}(a,0)$, $P_0(z^*)=a^*$, 
and $z(\tau,a,\epsilon)$ has continuous first derivatives with respect 
to $a, \epsilon$.  
\end{theorem}
$P_0$ is defined as a projection operator on the Banach space $S$
of continuous periodic functions of period T. If $f\in S$, write
$f=\mathrm{col}(g,h)$ where $g$ is a $p$ vector and $h$ is a $n-p$ 
vector, then
\[
P_0(f)=\mathrm{col}\left(T^{-1}\int_0^T g(t)\, dt,0 \right)
\]
So, $P_0$ brings an element $f$ in $S$ to a constant vector which
has the average values of $g$ over one period as the first $p$ 
components and zeros as the rest components. The equation satisfied 
by $z^*$ is different from (\ref{sys1}) by a constant vector. By 
a proper choice of the starting vector $a$, we may make this 
constant vector zero to obtain a solution for the system (\ref{sys1}). 
The mathematical statement is give by the following theorem.
\begin{theorem} \label{ther2}
Let $z(\tau,a,\epsilon)$ be the function given by the 
Theorem~\ref{ther1} for all $\|a\| \leq b <d, 
|\epsilon| \leq \epsilon_1$. If there exist an 
$\epsilon_2 \leq \epsilon_1$ and a continuous function 
$a(\epsilon)$ such that 
\begin{equation}
P_0Z(\tau,z(\tau,a(\epsilon),\epsilon),\epsilon)=0, 
\;\mbox{with }\|a(\epsilon)\| \leq b 
\mbox{   for } |\epsilon| \leq \epsilon_2 \label{cond1} 
\end{equation}
then $z(\tau,a(\epsilon),\epsilon)$ is a periodic solution 
of system (\ref{sys1}) for $\|\epsilon\| \leq \epsilon_2$. 
Conversely, if system (\ref{sys1}) has a periodic solution 
$\bar{z}(\tau,\epsilon)$, of period $T$, $\|\bar{z} (\tau,\epsilon)\| 
\leq d, |\epsilon| \leq \epsilon_2$, then $\bar{z} (\tau,\epsilon)
=z(\tau,a(\epsilon),\epsilon)$. 
\end{theorem}
Therefore, the existence of a continuous function $a(\epsilon)$ 
satisfying (\ref{cond1}) is a necessary and sufficient condition 
for the existence of a periodic solution of system (\ref{sys1}) 
of period $T$. As we do not know the exact functional form of 
the periodic solution, the condition (\ref{cond1}) could not be 
solved explicitly. But by using implicit function theorem, we 
can show that the substitution into (\ref{cond1}) of a proper
approximate function of $z(\tau,a,\epsilon)$ leads to the 
existence condition for periodic solutions. 
\begin{theorem} \label{ther3}
In the system (\ref{sys1}), let
\[
Z=\mathrm{col}(X,Y), \quad z=\mathrm{col}(x,y)
\]
where $X,x$ are $p$ vectors and define
\[
X_0(x,y,\epsilon)=\frac{1}{T}\int_0^T X(\tau,x,y,\epsilon) d\tau.
\]
If there is a $p$ vector $a_0, \|a_0\|<d $, such that
\begin{equation}
X_0(a_0,0,0)=0, \quad \det \left[\frac{\partial X_0(a_0,0,0)}{\partial x}\right]
\neq 0  \label{cond2}
\end{equation}
then there exists an $\epsilon_1>0$ and a periodic function 
$z(\tau,\epsilon), |\epsilon| \leq \epsilon_1$, of system 
(\ref{sys1}) of period $T$ with $z(\tau,0)=\mathrm{col}(a_0,0)$.
\end{theorem} 

If we need to determine other parameters as a function of $\epsilon$ in
practical applications, similar theorems could be derived. Specifically,
in the main text we consider the period $T$ as a function of $\epsilon$.
It is clear that theorem~\ref{ther3} applies if we suppose $T(\epsilon)$
is continuous in $\epsilon$ and bounded for $|\epsilon| \leq
\epsilon_1$. Furthermore, despite the use of the zeroth approximation in
the above theorem, the $n$th approximation could be used instead. If
{\em simple} (non-vanishing determinant) solutions to the determining
equations can be found for $\epsilon$ in the neighborhood of 0
then system (\ref{sys1}) has a periodic solution.

If the system which we are studying possesses certain symmetries, we can
prove the existence of particular symmetric solutions by a simplified
version of determining equations. Let us define the symmetry first.

\begin{defit}
Let $\dot{z}=f(\tau,z)$, where $z,f \in \mathbb{R}^n$, be a system of
differential equations. It is said to have the property $E$ with respect
to $Q$ if there exists a nonsingular matrix $Q$ such that
\[
Q^2=I \quad Qf(-\tau,Qz)=-f(\tau,z) \quad QP_0=P_0 Q 
\]
where $P_0$ is the projection operator defined before. \label{def1}
\end{defit}     

Under this symmetry assumption the following theorems apply:
\begin{theorem} \label{sym1}
Suppose $Q=\mathrm{diag}(Q_1,Q_2)$ where $Q_1$ is a 
$p \times p$ matrix. If system (\ref{sys1}) has property $E$ 
with respect to this $Q$ for all $\epsilon$. If $a,\|a\| \leq b,$ 
is a $p$ vector and $a^*=\mathrm{col}(a,0)$ is a $n$ vector, 
chosen in such a way that $Q a^*=a^*$, then the solution 
$z(\tau,a,\epsilon)$ satisfies the relation
\[
Qz(-\tau,a,\epsilon)=z(\tau,a,\epsilon)
\] 
and consequently,
\[
Z(-\tau,z(-\tau,a,\epsilon),\epsilon)=-Qz(\tau,z(\tau,a,\epsilon),\epsilon)
\]
\end{theorem}
\begin{theorem} 
\label{sym2}
If the $j$-th element of the diagonal of the matrix $Q_1$ in 
Theorem~\ref{sym1} is $+1$, then the $j$-th equation in the determining
equations is equal to zero for every vector $a^*$ in Theorem~\ref{sym1}.
\end{theorem}

The system (\ref{eq:4d}) derived here from the 1-$D$ CGLe has this
symmetry, so the number of determining equations can be reduced using
these two theorems.


\end{document}